\font\small=cmr10 scaled 1000               
%
%
\magnification=\magstep1
\baselineskip=11pt plus .1pt minus .1pt
\hsize=12.5truecm
\vsize=19.0truecm  
\hfuzz=5pt\vfuzz=5pt
\tolerance=1000
\overfullrule=0pt
\parskip=0pt
\abovedisplayskip=3 mm plus6pt minus 4pt
\belowdisplayskip=3 mm plus6pt minus 4pt
\abovedisplayshortskip=0mm plus6pt minus 2pt
\belowdisplayshortskip=2 mm plus4pt minus 4pt
\predisplaypenalty=0
\clubpenalty=10000
\widowpenalty=10000
\parindent=2em
%
%
\font\pgnumfont=cmr9
\font\headlinefont=cmti9
\font\titlefont=cmbx10
\font\authorfont=cmr10
\font\addressfont=cmti9
\font\datefont=cmr9
\font\sumfont=cmr9

\font\absfont=cmbx9
\font\secfont=cmr10
\font\subsecfont=cmti10
\font\subsubsecfont=cmr10
\font\figfont=cmr9
\font\figheadfont=cmbx9

\font\tabheadfont=cmbx9
\font\mainfont=cmr10
\font\petitrm=cmr9

%
%
%
\newtoks\TITLE \newtoks\AUTHOR \newtoks\ADDRESS \newtoks\SUMMARY
\newdimen\sumindent \sumindent=\parindent
\newtoks\KEYWORDS \newtoks\SUBMITTED \newtoks\ACCEPTED
\newtoks\SENDOFF
%

%
%
\newtoks\firstpage
\let\firstpage=Y
\newtoks\AUTHORHEAD \newtoks\ARTHEAD \newtoks\VOLUME \newtoks\PAGES
\if!\the\AUTHORHEAD!\AUTHORHEAD={\the\AUTHOR}\fi
\if!\the\ARTHEAD!\ARTHEAD={\the\TITLE}\fi
\footline={\hfil}
\headline={\ifodd\pageno\rightheadline \else\leftheadline\fi}
\def\leftheadline{\if Y\firstpage\firsthead\global\let\firstpage=N
  \else\lefthead\fi}
\def\rightheadline{\if Y\firstpage\firsthead\global\let\firstpage=N
  \else\righthead\fi}
\def\lefthead{\pgnumfont\number\pageno\hfil\headlinefont\the\AUTHORHEAD}
\def\righthead{\headlinefont\the\ARTHEAD\hfil\pgnumfont\number\pageno}
\def\firsthead{\headlinefont Baltic Astronomy,~vol.\the\VOLUME,
\the\PAGES,~\the\year .\hfil}
\voffset=2\baselineskip 
%

\newdimen\oldbaselineskip \oldbaselineskip=\baselineskip
\def\test#1{\newlinechar=`@\if!\the#1! \message{#1 not given@}\fi}%
\def\printheader{
  \parindent=0pt
  \null\vskip1.cm
  \test{\TITLE}
  \vbox{\baselineskip=15pt
    \titlefont\the\TITLE
    }
  \vskip8mm plus8mm
  \test{\AUTHOR}
  \authorfont\the\AUTHOR
  \vskip2mm
  \test{\ADDRESS}
  \addressfont\the\ADDRESS
  \vskip2mm
  \test{\SUBMITTED}
  \line{\datefont Received \the\SUBMITTED
    \if!\the\ACCEPTED!\else, accepted \the\ACCEPTED\fi.\hfill}
  \vskip4mm plus4mm
  \vbox{\leftskip=\sumindent\parindent=0pt
    \parskip=5pt
    \absfont Abstract.
    \test{\SUMMARY}
    \sumfont\the\SUMMARY\par
    \absfont Key words:
    \test{\KEYWORDS}
    \sumfont\the\KEYWORDS\par
    }
  \sumfont
  \if!\the\SENDOFF!\else\footnote{}{
 \the\SENDOFF}\fi
  \parindent=2em
  }
%
%
\newdimen\uppergap \newdimen\lowergap
\uppergap=5mm \lowergap=3mm
\newdimen\secind \newdimen\subsecind \newdimen\subsubsecind
\setbox0=\hbox{\secfont 9. }\secind=\wd0
\setbox0=\hbox{\subsecfont 9.9. }\subsecind=\wd0
\setbox0=\hbox{\subsubsecfont 9.9.9. }\subsubsecind=\wd0
\def\section#1{\goodbreak\par\vskip\uppergap
  \noindent\hangindent\secind\hangafter=1\secfont#1
  \vskip\lowergap\mainfont\par\nobreak}
\def\subsection#1{\goodbreak\par\vskip\uppergap
  \noindent\hangindent\subsecind\hangafter=1\subsecfont#1
  \vskip\lowergap\mainfont\par\nobreak}
\def\subsubsection#1{\goodbreak\par\vskip\uppergap
  \noindent\hangindent\subsubsecind\hangafter=1\subsubsecfont#1
  \vskip\lowergap\mainfont\par\nobreak}
%
%
%

%

%
\newdimen\tabind
\setbox0=\hbox{\tabheadfont Table 55.} \tabind=\wd0

%
%
\def\References{\vskip\uppergap
\line{\secfont REFERENCES\hfill}
  \vskip0.8\lowergap
 \petitrm
  }
\def\ref{\goodbreak
\hangindent12pt\hangafter=1
\noindent\ignorespaces}
\def\endref{\egroup}

\def\ref{\goodbreak
\hangindent12pt\hangafter=1
\noindent\ignorespaces}
\def\endref{\egroup}
%
%
\def\byebye{\egroup\par\vfill\supereject\end}
%
%

%
%

\def\utw{\smash{\rlap{\lower5pt\hbox{$\sim$}}}}
\def\udtw{\smash{\rlap{\lower6pt\hbox{$\approx$}}}}

\newdimen\free\newdimen\shift
\def\Entry#1#2#3{\par\goodbreak\smallskip%
  \setbox1=\vbox{\advance\hsize by-10mm\parindent=0pt
    \def\\{\par}%
    \it#1. \rm#2}
  \line{\box1\hfill#3}\smallskip
}%
\newdimen\savesize

\def\shiftfigure #1#2#3#4#5{
    \vbox to #2 { \ifodd #5 \rightskip#4 \else\leftskip#4 \fi
                  \null\vfil
                  \figheadfont Fig.~#1.\figfont #3
                  \medskip
                }
                          }

\year1996

\VOLUME{0}
\PAGES{00-00}
\pageno=1

\vskip -1.0 cm      
\noindent  
{\small  
Proc.~~{\it Internat.~Cooperation in Dissemination of Astronomical Data},} \break
\vskip -7 mm
{\small   
\centerline{July 3--5, 1996,~~St.-Petersburg, Russia}} 
\vskip -0.6 cm

\TITLE={WWW ACCESS~~TO~~RADIO MEASUREMENTS~~OF
CLUSTERS OF GALAXIES}
\AUTHOR={A.G.~Gubanov$^1$ and H.~Andernach$^2$}

\ARTHEAD={WWW access to galaxy cluster radio data} 
\AUTHORHEAD={Gubanov~~\&~~Andernach}    

\ADDRESS={$^1$Astronomical Institute, St.-Petersburg State University, 198904 Russia

$^2$IUE Observatory, Villafranca, Apdo.~50727, E--28080 Madrid, Spain}

\SUBMITTED={July 20, 1996}

\SUMMARY={Radio measurements of rich Abell clusters of galaxies have
been collected both from radio source catalogues and dedicated 
publications.
By gathering these data for every cluster, complemented by optical data,
an overall picture of each cluster may be obtained and 
samples may be drawn suitable for~~statistical~~studies.
         }

\KEYWORDS={
clusters: general; clusters: radio sources; databases
          }

\printheader

\vskip 5 mm

Radio sources in clusters of galaxies provide an important sample to study
the physics of the intracluster medium, its influence on radio sources, and
the overall evolution of extragalactic radio sources.
Numerous cluster observations are available in different spectral ranges
but the relevant data are spread over hundreds of publications
and are heterogeneous with respect to quality, sensitivity, angular resolution,
availability of an optical identification, etc.. They suffer from different
selection effects, and only part of them is available in electronic form.
This makes it difficult to draw correct astrophysical conclusions from 
a combination of different radio data.

To overcome these obstacles and to allow more efficient selection of source 
samples suitable to answer astrophysical questions about cluster 
radio emission, we are developing a database of radio sources in clusters of 
galaxies. It is based on the CGI interface and HTML forms and accessible 
on the WWW server of the Astronomical Institute of Sankt-Petersburg 
University (AISPbU) under URL \hskip 6 cm \break
\vskip -3 mm
\hskip 5 mm {\tt http://www.aispbu.spb.su/WWW/Clusters.html}

The individual source measurements were extracted partly from the Radio 
Astronomical Catalogues (RAC) database of AISPbU (Gubanov \& Titov 1996)
and partly from many publications dedicated to cluster radio emission. 
Dozens of published source tables were converted by us into electronic form 
for the first time, using a page scanner with ``Optical Character 
Recognition'' software (OCR) and special proof-reading tools~~~
(cf. Andernach et al. 1996, these proceedings). 

We currently include in our service radio and optical data for all 4076
rich clusters in the ACO catalogue (Abell et al.~1989).
Inclusion of X-ray information and properties of the brightest cluster 
galaxies is planned.
We try to realise the following basic ideas in this service:
\item {--} gather in one place all major information for a given cluster;
\item {--} create an overall picture of the cluster based on measurements;
\item {--} give a clear graphic presentation of different cluster data;
\item {--} allow users to provide corrections and supply new data;
\item {--} provide tools for statistical work with cluster data.

At present we have built the necessary tools for all except the last
item.
The URL given above allows to extract and display either radio or optical 
data for a user-specified cluster. 

The page with {\bf radio data} begins with comments on certain radio
sources, e.g.~on cluster membership of sources, ambiguities of the
interpretation of data, etc.   This is
followed by a section on individual radio galaxies in the cluster, 
including the optical identification and radio spectral data,
sorted by observing frequency and converted to a common absolute
flux scale where possible.
The last block of the page gives the data {\it as published} from all 
relevant references for radio sources within a projected distance of 
one Abell radius from their centres (R$_{\rm a}$ = 3 Mpc for H$_{\circ}$=50 
kms$^{-1}$Mpc$^{-1}$).
More detail on these references is available and in future we plan to provide
hyperlinks to abstract services or to electronic journals.
Further links allow to extract entries from the preliminary source list
of the NRAO VLA Sky Survey (NVSS, Condon et al. 1996) 
and to display the Digitized Sky Survey image of a 500$\times$500 kpc$^2$ 
region around a given cluster radio galaxy.

The page with {\bf optical data} lists the basic optical parameters of the
cluster, as well as    
published notes and relevant references. We began to provide links
to optical object catalogs from APM (Irwin et al. 1994) which
may be displayed as PostScript finding charts, currently limited to a maximum
field of 30$'\times$30$'$, i.e.~less than 1 R$_{\rm a}$ for nearby clusters. 
A more promising tool
are Tcl applets, which permit to run application programs on the 
user's local computer within the WWW connection.
Applets allow to show the relevant parameters 
of the displayed objects simply by positioning the cursor on them. 
They require less disk space than PostScript
and allow the user to control the design of the charts. 
We plan to use Tcl applets
later to link the optical images of objects in the cluster
with any relevant information available on them, like radio parameters, 
spectra, redshifts, etc. 
We currently use Tcl applets either with the SurfIt! browser
(Titov 1996, these proceedings) or with the netscape browser
using the ``Simple windowing shell'' ({\tt wish}) program. These should also
directly work under the netscape browser soon.

We anticipate that many users are more familiar with specific sections 
of data than we are. To improve the quality and completeness of the 
cluster database it is thus important that the user may provide 
corrections or additions to the data in an as comfortable fashion
as possible.  Rather than having to compose a separate email message, 
the user may simply prepare revised versions of the data ``on the fly'' 
without leaving their browsers, using {\tt xedit}
(under X-windows) or the {\tt vi} editor (otherwise).
Revisions are folded into the database upon our check and approval.  
Depending on first experience with input from users we
can arrange for other editors if needed.

The database is being updated permanently with
new cluster radio and optical measurements, e.g. 
from the NRAO VLA Sky Surveys NVSS and FIRST (Becker et al. 1995),
or with optical objects from APM, COSMOS (Drinkwater et al. 1995)
and APS (cf. Odewahn 1995). 

Our service is intended as an open and collaborative enterprise
to the benefit of cluster research.  We welcome users to join our efforts,
to tell us their preferences for data to be included and 
to contribute new data sets, especially
those not previously available in electronic form.

\vskip 2 mm
This work is supported by the RFBR grant N 94-07-20441.

\References

\ref
Abell G.O., Corwin Jr. H.G. \& Olowin R.P. 1989, ApJS 70, 1 
\ref 
Becker R.H., White R.L. \& Helfand D.J. 1995, ApJ, 450, 559
\ref
Condon J.J. et al. 1996, AJ, to be submitted
\ref
Drinkwater M.J., Barnes D.G. \& Ellison S.L. 1995, Proc. ASA, 12, 248
\ref
Gubanov A.G. \& Titov V.B. 1996, Vestnik St.-Petersburg Univ., 
Ser. 1, Vyp. 3 (N 15), in press
\ref
Irwin M., Maddox S., \& McMahon R. 1994, Spectrum, Royal Obs. 2, 14
\ref
Odewahn S. C. 1995, PASP 107, 770
\bye